\documentclass[aps,showpacs,showkeys,nofootinbib,floatfix]{revtex4}

\input psfig.sty
\usepackage{amssymb}
\usepackage{amsmath}
\usepackage{graphicx}

\begin{document}

\title{Milliarcsecond compact structure of radio quasars and the geometry of the Universe}

\author{Shuo Cao$^{1\ast}$}
\author{Jingzhao Qi$^2$}
\author{Marek Biesiada$^{3\dagger}$}
\author{Xiaogang Zheng$^1$}
\author{Tengpeng Xu$^1$}
\author{Yu Pan$^{4}$}
\author{Zong-Hong Zhu$^{1\ast}$}

\affiliation{$1.$ Department of Astronomy, Beijing Normal
University, Beijing 100875, China; \emph{caoshuo@bnu.edu.cn; zhuzh@bnu.edu.cn}\\
$2.$ Department of Physics, College of Sciences, Northeastern
University, Shenyang 110004, China; \\
$3.$ Department of Astrophysics and Cosmology, Institute of Physics,
University of Silesia, ul. 75 Pu{\l}ku Piechoty 1, 41-500 Chorz{\'o}w, Poland;
\emph{marek.biesiada@us.edu.pl}\\
$4.$ College of Science, Chongqing University of Posts and
Telecommunications, Chongqing 400065, China }

\begin{abstract}

In this paper, by using the recently compiled set of 120
intermediate-luminosity quasars (ILQSO) observed in a
single-frequency VLBI survey, we propose an improved
model-independent method to probe cosmic curvature parameter
$\Omega_k$ and make the first measurement of the cosmic curvature
referring to a distant past, with redshifts up to $z\sim 3.0$.
Compared with other methods, the proposed one involving the quasar
data achieves constraints with higher precision in this redshift
range. More importantly, our results indicate that the measured
$\Omega_k$ is in good agreement with zero cosmic curvature, implying
that there is no significant deviation from a flat Universe.
Finally, we investigate the possibility of testing $\Omega_k$ with a
much higher accuracy using quasars observed in the future VLBI
surveys. It is shown that our method could provide a reliable and
tight constraint on the prior $\Omega_k$ and one can expect the zero
cosmic curvature to be established at the precision of
$\Delta\Omega_k\sim 10^{-2}$ with 250 well-observed radio quasars.

\end{abstract}


\maketitle

\section{Introduction}

The question of whether the Universe is spatially open, flat or
closed is one of the most fundamental issues influencing our
understanding of the Universe, its structure, evolution and matter
budget. Possibilities that cosmic curvature deviates from the zero
value may have far-reaching consequences for our knowledge of
fundamental problems like validity of the
Friedman-Lema\^{\i}tre-Robertson-Walker (FLRW) approximation
\citep{Ferrer06,Ferrer09} or the history of the Universe (inflation
theory, observed late-time accelerated expansion, reconstruction of
the equation of state of dark energy)
\citep{2007JCAP...08..011C,2008JCAP...12..008V}.

Although the zero value of the cosmic curvature is currently
supported by many astrophysical probes -- in particular very
strongly by the first acoustic peak location in the pattern of
anisotropies of the  the Cosmic Microwave Background Radiation
(CMBR) -- two
issues should be reminded. First, CMBR measurements 
e.g. the latest 
\textit{Planck} 2015 results  \citep{Planck1}, are not direct, but
strongly depend on the pre-assumed cosmological model (the standard
$\Lambda$CDM model). Second, alternative methods of deriving the
spatial
curvature from popular probes 
\citep{2007JCAP...08..011C,2008PhRvL.101a1301C,2010PhRvD..81h3537S,Li16,Wei16},
focus on the luminosity distance $D_{L}(z)$ using SN Ia as standard
candles at lower redshifts ($z\sim 1.40$), combined with the
measurements of the Hubble parameter $H(z)$ using passively evolving
galaxies and baryon acoustic oscillation (BAO) as cosmic
chronometers. 
These results showed no evidence for the deviation from flatness
(see also~\citet{2011arXiv1102.4485M,2014PhRvD..90b3012S}).

Using SN Ia data to determine the cosmic curvature one should
remember that the nuisance parameters characterizing SN Ia
light-curves introduce considerable uncertainty to the final
determination of $\Omega_k$ \citep{Li16,Wei16}. More important,
however, is that with SN Ia we are able to probe the lower redshift
range $z\leq 1.40$ \citep{Amanullah10,JLA2014}, while the CMBR
measurements refer to a much higher redshift $z\sim 1000$
\citep{Planck1}. The so-called ``redshift desert'' problem still
remains challenging with respect to the exploration of the cosmic
curvature. Therefore, attempting to, at least partly, bridge this
``desert''  we perform an improved model-independent test of cosmic
curvature with the QSO data extending our investigation to $z\sim
3.0$.

\section{Observational data}

The advances in cosmology over recent decades have been accompanied
by intensive searches for reliable standard rulers at higher
redshifts. One class of promising candidates in this context are
ultra-compact structures in radio quasars that can be observed up to
high redshifts, with milliarcsecond angular sizes measured by the
very-long-baseline interferometry (VLBI)
\citep{Kellermann93,Gurvits94}. The observed angular size
$\theta(z)$ of the compact structure in each quasar is given by
\citep{Sandage88}
\begin{equation}
\theta(z)= \frac{l_m}{D_A(z)} \label{theta}
\end{equation}
where $l_m$ is the intrinsic metric linear size of the source and
$D_A$ is the angular diameter distance. Ultracompact objects are
identified as cases in which the jets are moving relativistically
and are close to the line of sight, when Doppler boosting allows
just that component which is moving towards the observer to be
observed. To be more specific, at any given frequency (and we use a
single frequency data), the core is believed to be located in the
region of the jet where the optical depth is $\tau = 1$, with
synchrotron self-absorption being dominant source of opacity.
However, the problem is that, in general, linear sizes $l_m$ of
compact radio sources might not be constant. The precise value of
$l_m$ might depend both on redshifts and the intrinsic properties of
the source, i.e., its luminosity. \citet{Gurvits94} attempted to
estimate how much evolution was actually occurring as a function of
redshift and to what degree this affected the optimization of the
model parameters. \citet{Gurvits99} modeled the luminosity
dependence of the compact source size as $l_m=lL^\beta(1+z)^n$, with
two parameters $\beta$ and $n$ parameterizing the possible
dependence of the intrinsic size on the luminosity and the redshift.
The $\beta$ parameter, which captures the dependence of the linear
size on source luminosity, is highly dependent on the physics of
compact radio emitting regions \citep{Jackson06}. Besides
cosmological evolution of the linear size with redshift, the
parameter $n$ may also characterize the dependence of the linear
size on image blurring due to scattering in the propagation medium
\footnote{Considering that all VLBI images for our sample were
observed at a frequency of 2.29 GHz, this effect is not important in
the present analysis \citep{Gurvits99,Cao15}.} \citep{Gurvits99}.
One should also clarify the dependence of the angular size on the
intrinsic emitting frequency $\nu_e$. Namely, the data are obtained
at a single receiving frequency $\nu_r$, which corresponds to an
emitting frequency that scales up as $(1+z)$: $\nu_e=(1+z)\nu_r$,
thus masking possible cosmological effects. According to the
core-jet structure observed in typical VLBI images
\citep{Pushkarev12}, the core is believed to be the base of the jet, rather than the nucleus
 \citep{Blandford79}. However,
\citet{Dabrowski95} found that as $z$ increases, a larger Doppler
boost factor {\cal D} is required for a flux-limited radio sample.
On the other hand, \citet{Jackson04} provided a very thorough
discussion of this issue, in the context of compact sources like
those used in this paper. His conclusion was that an approximately
fixed ratio ${\cal D}/(1 + z)$ can be expected, which will generate
an approximately fixed rest-frame emitted frequency $(1 +
z)\nu_r/{\cal D}$. In other words, statistically the cosmological
redshift effect is roughly cancelled out by the Doppler boost
\citep{Jackson04}.

Our study is based on the sub-sample identifed and calibrated in
\citep{Cao15,Cao17a}, so a brief description of this sample is in
order. The original source of the data was a well-known 2.29 GHz
VLBI survey undertaken by \citet{Preston85} (hereafter called P85).
By employing a world-wide array of dishes forming an interferometric
system with an effective baseline of about $8\times 10^7$
wavelengths, this survey succeeded in detecting interference fringes
from 917 radio sources out of a list of 1398 candidates selected
mainly from the Parkes survey \citep{Bolton79}. This work was
extended further by \citet{Jackson06}, who updated the P85 sample
with respect to redshift, to include a total of 613 objects with
redshifts $0.0035\leq z\leq 3.787$. The full listing is available in
electronic form \footnote{http://nrl.northumbria.ac.uk/13109/},
including source coordinates, redshift, angular size, uncertainty in
the latter, and total flux density. Later on, \citet{Cao17a} divided
the Jackson's sample into different sub-samples, according to their
optical counterparts and luminosity: low, intermediate, and
high-luminosity quasars. Luminosity selection as well as $D_A(z)$
assessments necessary for building the sample were performed without
pre-assuming a cosmological model but basing on the $D_A(z)$
reconstruction from $H(z)$ data obtained from cosmic chronometers
\citep{Jimenez}. It was found that 120 quasars with
intermediate-luminosities (ILQSO) show negligible dependence on both
redshifts $z$ and intrinsic luminosity $L$ ($|n|\simeq 10^{-3}$,
$|\beta|\simeq 10^{-4}$), which means they meet the requirements
expected from standard rulers. The redshift of these ILQSO ranges
between $z=0.462$ and $z=2.73$. Let us remind that the angular sizes
$\theta$ of these standard rulers were estimated from the ratio of
total and correlated flux densities measured with radio
interferometers ($\Gamma=S_c/S_t$), i.e., the visibility modulus
$\Gamma$ defines a characteristic angular size
\begin{equation}
\theta={2\sqrt{-\ln\Gamma \ln 2} \over \pi B} \label{thetaG}
\end{equation}
where $B$ is the interferometer baseline measured in wavelengths.
Subsequently, \citet{Cao17b} used an improved
cosmological-model-independent method to calibrate the linear sizes
of ILQSO as $l_m=11.03\pm0.25$ pc at 2.29 GHz. Cosmological
application of this data set \citep{Cao17b} resulted with stringent
constraints on both the matter density $\Omega_m$ and the Hubble
constant $H_0$, in a very good agreement with recent \textit{Planck}
results. The constraining power of the quasar data was also studied
in different variable gravity models
\citep{Li17,Ma17,Qi17,Xu17,Zheng17}.

An issue which needs clarification is the achievable resolution with
VLBI, i.e., whether the quasars are truly resolved in the VLBI
survey data. Resolution criterion considering the visibility
distribution corresponding to the VLBI core, proposed in
\citet{Lobanov05} and extensively discussed in the literature
\citep{Kovalev05,Jackson12}, states that minimum resolvable size of
a Gaussian component fitted to naturally weighted VLBI data is:
\begin{equation}
\theta_{\mathrm{lim}} =b\left[{4\ln 2 \over \pi}\,\ln\left({SNR
\over SNR-1}\right)\right] ^{1/2},
\end{equation}
where $b$ is the half-power beam width taken as half of the fringe
spacing, $b=(2B)^{-1}$ \citep{Jackson12}. The signal--to--noise
ratio is defined as $SNR=S_{core}/\sigma_{core}$, where
$\sigma_{core}$ is the rms noise level in the area of the image
occupied by the core component, which can be measured from the
residual pixel value in the core component convolved with the
synthesized beam \citep{Kovalev05}. With two representative values
of $SNR$ mentioned in \citet{Preston85}, which account for a
systematic error of about 10\% and an absolute random error of 0.02
Jy in the correlated flux density, the corresponding resolution is
$\theta_{\mathrm{lim}}\sim 0.20$ mas \citep{Jackson12}, which can be
significantly smaller than the Rayleigh limit of $\sim 2.6$ mas with
a baseline $B=8\times 10^7$. On the other hand, considering the size
of the VLBI synthesized beam of the P85 survey, there are large
uncertainties for sources with angular sizes smaller than the size
of the resolving beam. However, the mean angular size of 120 quasars
used in this analysis ($\sim$1.45 mas), is greater than the
half-power beam width taken as half of the fringe spacing
($\sim$1.29 mas). One should emphasize that, in spite of the above
justifications, one can still be concerned about the lack of
adequate angular resolution (or lack of adequate coverage of spatial
frequencies). Namely, the source size characterization offered by
Eq.~(2) is based on just two uv-points and necessitates use of the
simplest source structure model, the efficiency of which acing as a
standard ruler still needs to be checked with a larger quasar sample
from future VLBI observations based on better uv-coverage. In this
paper we use the dataset published in \citet{Cao17b} fitting the
calibration parameter $l_m$ simultaneously with the cosmic curvature
parameter $\Omega_k$. As we will see the best fitted $l_m$ parameter
is robustly of $\sim 11$ pc scale.

Even though we do not fully understand physical reasons behind
ILQSOs behaving as standard rulers, yet there is some piece of
circumstantial evidence consistent with this assumption. In the
conical jet model proposed by \citet{Blandford79}, the unresolved
compact core is identified with the innermost, optically thick
region of the approaching jet. Current theoretical models strongly
indicate that Active Galactic Nuclei (AGNs) are powered by accretion
onto massive black holes (BH) \citep{Meier09}. The morphology and
kinematics of compact structure in radio quasars and AGN could be
strongly dependent on the nature of the ``central engine'',
including the mass of central black hole and the accretion rate,
etc. \citep{Gurvits99}. Therefore, the central region may be
``standard'' if these parameters are confined within restricted
ranges for specific quasars. Fortunately, there exists compelling
evidence indicating a correlation between black hole mass $M_{BH}$
and radio luminosity $L_R$ \citep{Lacy01,Jarvis02}. The direct link
between radio loudness and the black hole mass has been reported by
several authors \citep{Laor00,McLure04,Chiaberge11}, who found that
radio-loud objects are always associated with massive black holes.
Using black hole mass estimated from the full width at half-maximum
(FWHM) of the H$\beta$ line, a joint ADD analysis with the quasars
from the FIRST Bright Quasar Survey (FBQS) the Palomar-Green survey
(PG) quantitatively revealed the positive dependence of radio
luminosity on $M_{BH}$, as well as the accretion rate relative to
the Eddington limit (the ratio of bolometric luminosity to the
Eddington luminosity, $L_B/L_{Edd}\sim 0.1$):
\begin{equation}
\log{ L_R} = a \log{ M_{BH}} +  b \log{ L_B/L_{Edd}} +c
\end{equation}
Considering the realistic Doppler boosting correction
\citep{Jarvis02}, we reanalyzed the radio-loud FBQS and PG
sub-sample with flat spectrum ($\alpha<0.5$) and found our quasar
sample covering the luminosity of $10^{27} W/Hz<L_R<10^{28} W/Hz$
corresponds to the black hole mass ranging from $10^{8.5} M_\odot$
to $10^{9}M_\odot$. A new analysis of the relation between black
hole mass and radio luminosity in flat-spectrum quasars (FSQs) was
done in \citet{Jarvis02}, who investigated the black hole masses of
a sample of flat-spectrum radio-loud quasars from the Parkes
Half-Jansky Flat Spectrum sample of \citep{Drinkwater97}. Again,
following the application of a realistic Doppler boosting
correction, the mass of central black hole is confined to a
restricted range ($10^{8.5} M_\odot$ to $10^{9}M_\odot$) for our
quasar sample with intermediate luminosities ($10^{27}
W/Hz<L_R<10^{28} W/Hz$). More recent works focusing on up-to-date
mass estimates of black holes \citep{Chiaberge11}, which explicitly
took into account the effects of radiation pressure, concluded that
there are basically no radio-loud AGNs with $M_{BH}<10^{8} M_\odot$.
Therefore, the above studies suggest that, in order to produce a
powerful radio-loud quasar ($10^{27} W/Hz<L_R<10^{28} W/Hz$), a
large black hole mass ($10^{8.5} M_\odot$ to $10^{9}M_\odot$) with a
BH radius of influence ($\sim$11 pc) is a required element. Of
course, our treatment of the correlation between the radio
luminosity of quasars and the influence radius of the central black
hole should be considered with due caution, as the specific relation
between the mass of black hole and its influence radius still needs
to be justified quantitatively. Nevertheless these arguments points
towards the physical condition (central black hole mass limits)
under which intermediate-luminosity quasars could serve as
cosmological standard rulers.

Another set of data we use comprises the Hubble parameters $H(z)$
(i.e. the expansion rates at different redshifts). For this purpose
we use the latest 41 $H(z)$ data points obtained from the derivative
of redshift with respect to cosmic time obtained in two approaches.
One part of $H(z)$ is inferred from 31 passively evolving galaxies
\citep{2003ApJ...593..622J, 2005PhRvD..71l3001S, 2010JCAP...02..008S, 2012MNRAS.426..226C, 2014RAA....14.1221Z, Moresco15}
and the other part is derived from 10 radial baryon acoustic
oscillation (BAO) measurements
\citep{2009MNRAS.399.1663G,Blake12,Xu13,Samushia14,Busca2013,2014JCAP...05..027F,Delubac15}.
The redshift of the Hubble parameter data ranges from $z=0.09$ to
$z=2.34$, which makes $H(z)$ an efficient cosmological probe to
directly constrain cosmological parameters
\citep{Cao11a,Cao11b,Cao13,Chen15}, as well as perform fits on
cosmological parameters based on the two-point diagnostics
\citep{Ding15,Zheng16,Qi18,Zheng18}.

\begin{figure*}
\centering
\includegraphics[scale=0.35]{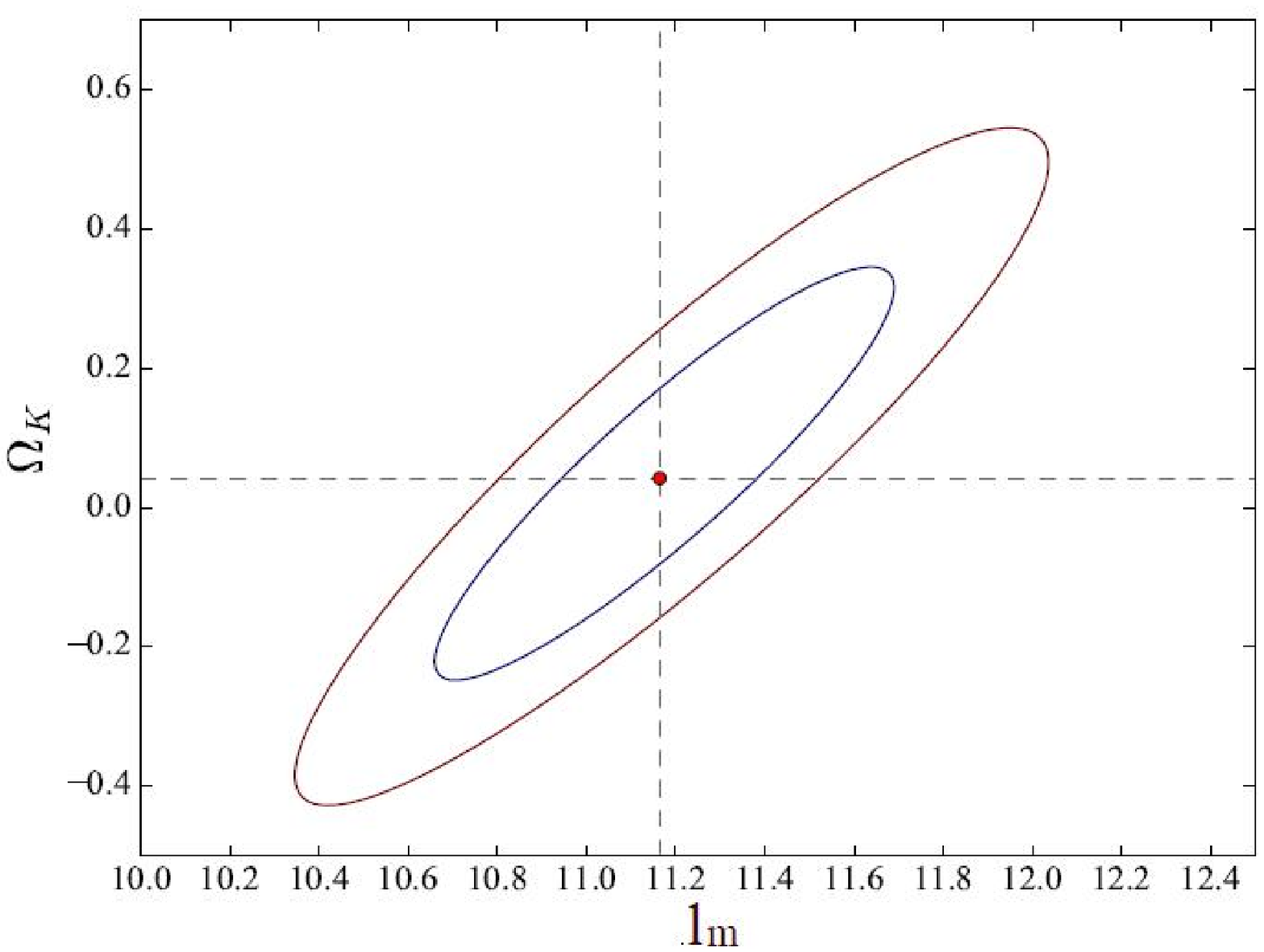} \includegraphics[scale=0.35]{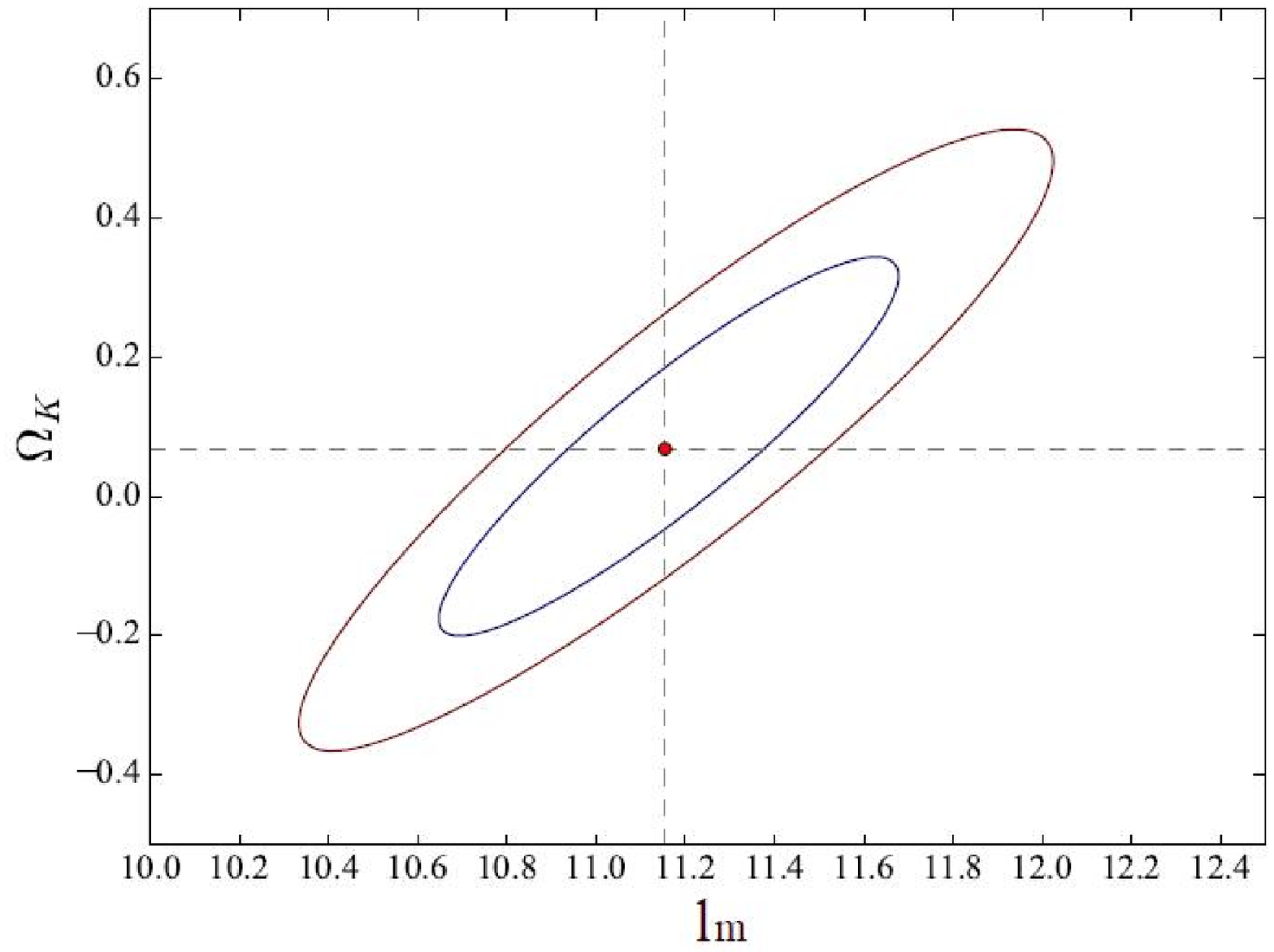}
\includegraphics[scale=0.35]{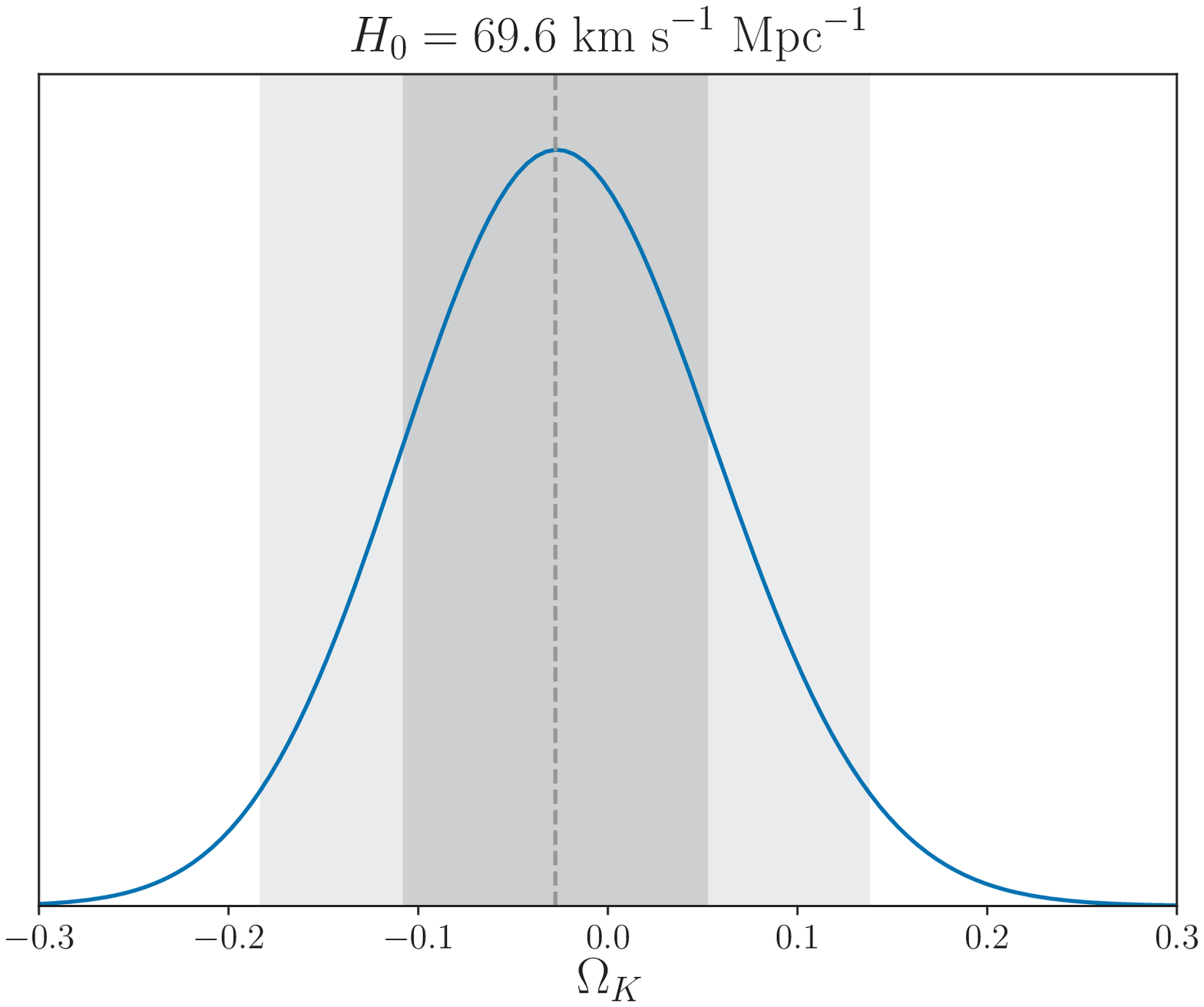} \includegraphics[scale=0.35]{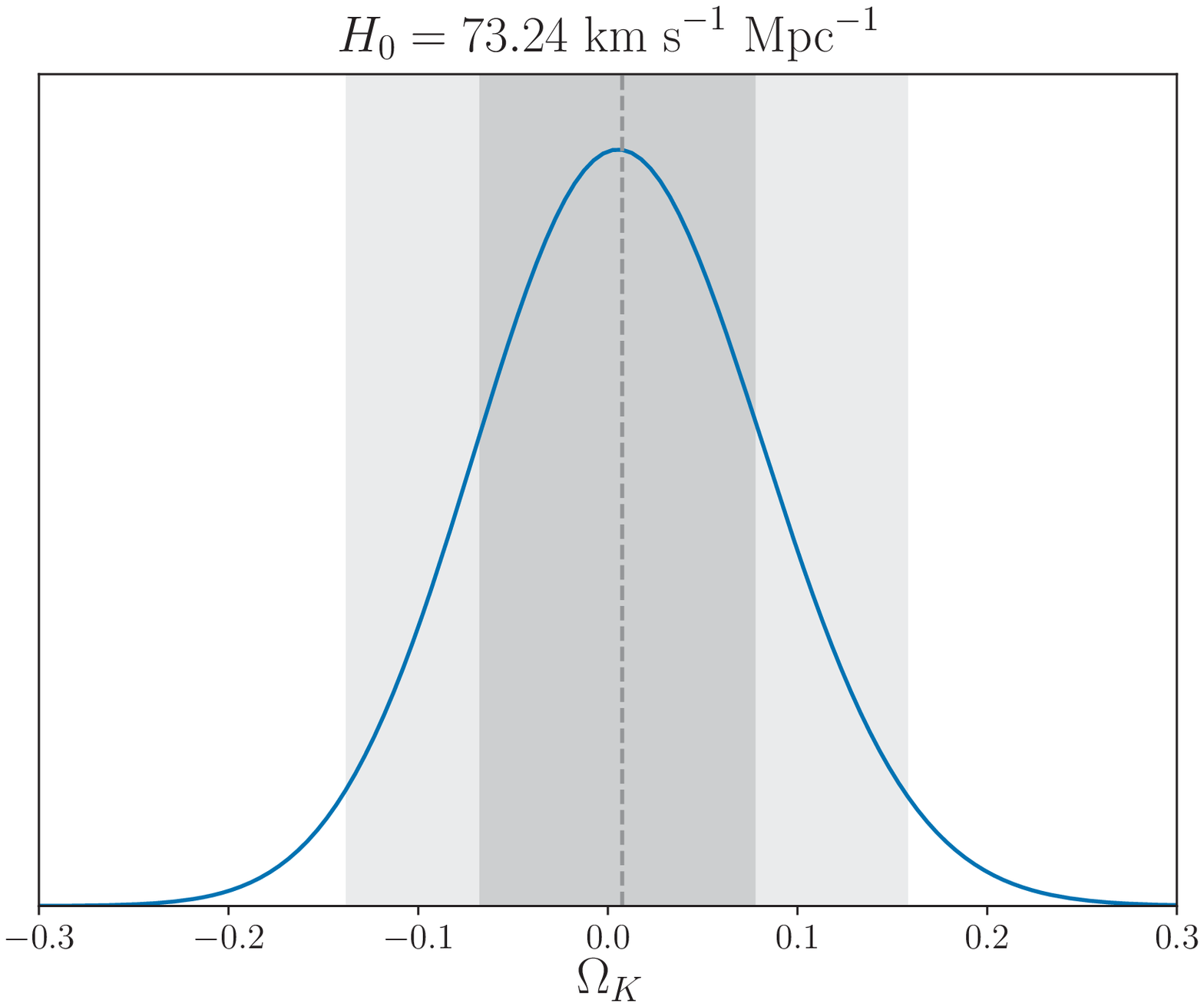}
\caption{ Results from a single-frequency VLBI observations of
ILQSOs: $1\sigma$ and $2\sigma$ constraint contours for $\Omega_{k}$
and $l_m$ (upper panel), and constraints on the cosmic curvature
with corrected linear size of compact quasars derived in a
cosmological model independent way (lower panel), corresponding to
two priors on the Hubble constant.}\label{fig1}
\end{figure*}

\section{Methodology}

The $H(z)$ data can be used to reconstruct the proper distance
\begin{equation}
D_{P}(z)= c \int^{z}_{0}\frac{dz'}{H(z')}
\end{equation}
and consequently also the angular diameter distance $D^{H}_{A}$ via
\begin{equation}\label{HSFR}
D^{H}_{A}(z) = \left\lbrace \begin{array}{lll} \frac{c}{H_{0}(1+z)}\frac{1}{\sqrt{|\Omega_{k}|}}\sinh\left[\sqrt{|\Omega_{k}|}D_{P}(z)\frac{H_{0}}{c}\right]\\
                                         \frac{1}{1+z}D_{P}(z)~~~~~~~~~~~~~~~~~~~~~~~~~~~~~~~~~~~\; \\
                                         \frac{c}{H_{0}(1+z)}\frac{1}{\sqrt{|\Omega_{k}|}}\sin\left[\sqrt{|\Omega_{k}|}D_{P}(z)\frac{H_{0}}{c}\right]\
\end{array} \right.
\end{equation}
for $\Omega_{k}>0$, $\Omega_{k}=0$ and $\Omega_{k}<0$, respectively.
Let us stress that this can be done without assuming any particular
model like the $\Lambda$CDM. Superscript ``H'' emphasizes this
issue. Model-independent reconstruction can be performed using the
Gaussian processes (GP). Technical details of this method as well as
description of the associated Python package can be found in
\citet{2012JCAP...06..036S}. Then one can define
$\theta_{H}(\Omega_{k};\,z)$ as the reconstructed angular-size of
the compact structure in radio quasars from the $H(z)$ data:
$\theta_{H}(\Omega_{k};\,z)= l_m/D^{H}_{A}(\Omega_{k};\,z)$.

As can be seen in Eq.(\ref{HSFR}), the spatial curvature
$\Omega_{k}$ directly enters the reconstructed theoretical
angular-size $\theta_{H}(\Omega_{k};\,z)$ of the compact structure
in radio quasars. This makes it possible to investigate $\Omega_{k}$
by confronting $\theta_{H}(\Omega_{k};\,z)$ with the observed value
of $\theta_{QSO}$. More important is that, as compared with the SN
Ia data extensively used in the literature \citep{Li16,Wei16}, there
is only one nuisance parameter $l_m$ in the distance estimate of
quasars, which should be fitted simultaneously with the cosmic
curvature parameter $\Omega_k$. In this work, we firstly take $l_m$
as a free parameter and justify to which extent the inferred
curvature depends on it. Then, a different
cosmological-model-independent method will be applied to determine
the linear size of this standard rod, which will help us to obtain a
cosmological - model - independent constraint on the cosmic
curvature $\Omega_k$. In order to explore the influence of the
Hubble constant on the reconstruction and then on the test of the
curvature parameter \citep{Wei16}, two recent measurements of
$H_{0}=69.6\pm0.7$ km $\rm s^{-1}$ $\rm Mpc^{-1}$ with $1\%$
uncertainty \citep{2014ApJ...794..135B} and $H_{0}=73.24\pm1.74$ km
$\rm s^{-1}$ $\rm Mpc^{-1}$ with $2.4\%$ uncertainty
\citep{2016ApJ...826...56R} have been used for distance estimation
in our analysis.

\begin{table}
\caption{\label{tabresult} Best-fit values and related $1\sigma$
uncertainties of the cosmic curvature $\Omega_{k}$ and the QSO
calibration parameter $l_m$, derived from ILQSO data.
$H_{0}(I)$ and $H_{0}(II)$ correspond to 
different priors of $H_{0}=69.6\pm0.7$ km $\rm s^{-1}$ $\rm
Mpc^{-1}$ \citep{2014ApJ...794..135B} and $H_{0}=73.24\pm1.74$ km
$\rm s^{-1}$ $\rm Mpc^{-1}$ \citep{2016ApJ...826...56R},
respectively.}
\begin{center}
\begin{tabular}{l|l|l}\hline\hline
QSO ($H_0$ prior) & Cosmic curvature
($\Omega_k$)  & Linear size ($l_m$) \hspace{4mm} \\
\hline
$H_{0}$(I) &  $\Omega_k=0.0\pm0.3$  & $l_m=11.2\pm0.5$ pc \\
                  &  $\Omega_k=0.0\pm0.1$ & $l_m=11.0\pm0.4$ pc \\

\hline
$H_{0}$(II) &  $\Omega_k=0.1\pm0.3$ & $l_m=11.2\pm0.5$ pc \\
                  &  $\Omega_k=0.0\pm0.1$ & $l_m=11.0\pm0.4$ pc \\
\hline\hline
\end{tabular}
\end{center}
\end{table}

\section{Results and discussion}

We determine the cosmic curvature $\Omega_k$ by minimizing the
$\chi^{2}$ objective function
\begin{equation}
\chi^{2}(l_m\;\Omega_{k})
=\sum_{i=1}^{120}\frac{[\theta_{H}(z_i;l_m,\Omega_{k})-\theta_{QSO}(z_i)]^2}{\sigma_{\theta}(z_i)^2}\;,
\end{equation}
with
$\sigma_{\theta}^2=\sigma^2_{\theta_{H}}+(\sigma^{sta}_{QSO})^2 + (\sigma^{sys}_{QSO})^2$,
where $\sigma^{sta}_{QSO}$ is the observational statistical
uncertainty for the \textit{i}th data point in the sample. Note that
the random uncertainty in correlated flux density is $\sim 0.02$ Jy,
while the corresponding random uncertainties in total flux-density
measurements typically range from 0.03 to 0.3 Jy \citep{Preston85}.
Moreover, we have added $10\%$ systematic uncertainties
($\sigma^{sys}_{QSO}$) in the observed angular sizes, which is
equivalent to an additional 10\% uncertainty accounting
for the intrinsic spread in the linear size \citep{Cao17b}. In order
to explore the influence of the Hubble constant on the inferred
curvature parameter, two priors of $H_0$ are taken into account in
our study. Results are shown in Fig.~\ref{fig1} and summarized in
Table 1.

To start with, by applying the above mentioned
$\chi^{2}$-minimization procedure, we obtain $1\sigma$, $2\sigma$
contours for the joint distributions of $\Omega_k$ and $l_m$ in
Fig.~\ref{fig1}. From this figure one can see the covariance between
the cosmic curvature and the intrinsic linear size of quasars. With
the prior of $H_{0}=69.6\pm0.7$ km $\rm s^{-1}$ $\rm Mpc^{-1}$, the
best-fitted parameters with corresponding $1\sigma$ uncertainties
are $\Omega_k=0.0\pm0.3$ and $l_m = 11.2\pm0.5$ pc. With the prior
of $H_{0}=73.24\pm1.74$ km $\rm s^{-1}$ $\rm Mpc^{-1}$, the
best-fitted parameters are $\Omega_k=0.1\pm0.3$ and $l_m=11.2\pm0.5$
pc, respectively. One can easily see that, for both $H_{0}$ priors,
estimation of the spatial curvature using quasars is fully
compatible with flat Universe at the current level of observational
precision.

\begin{figure*}
\centering
\includegraphics[scale=0.45]{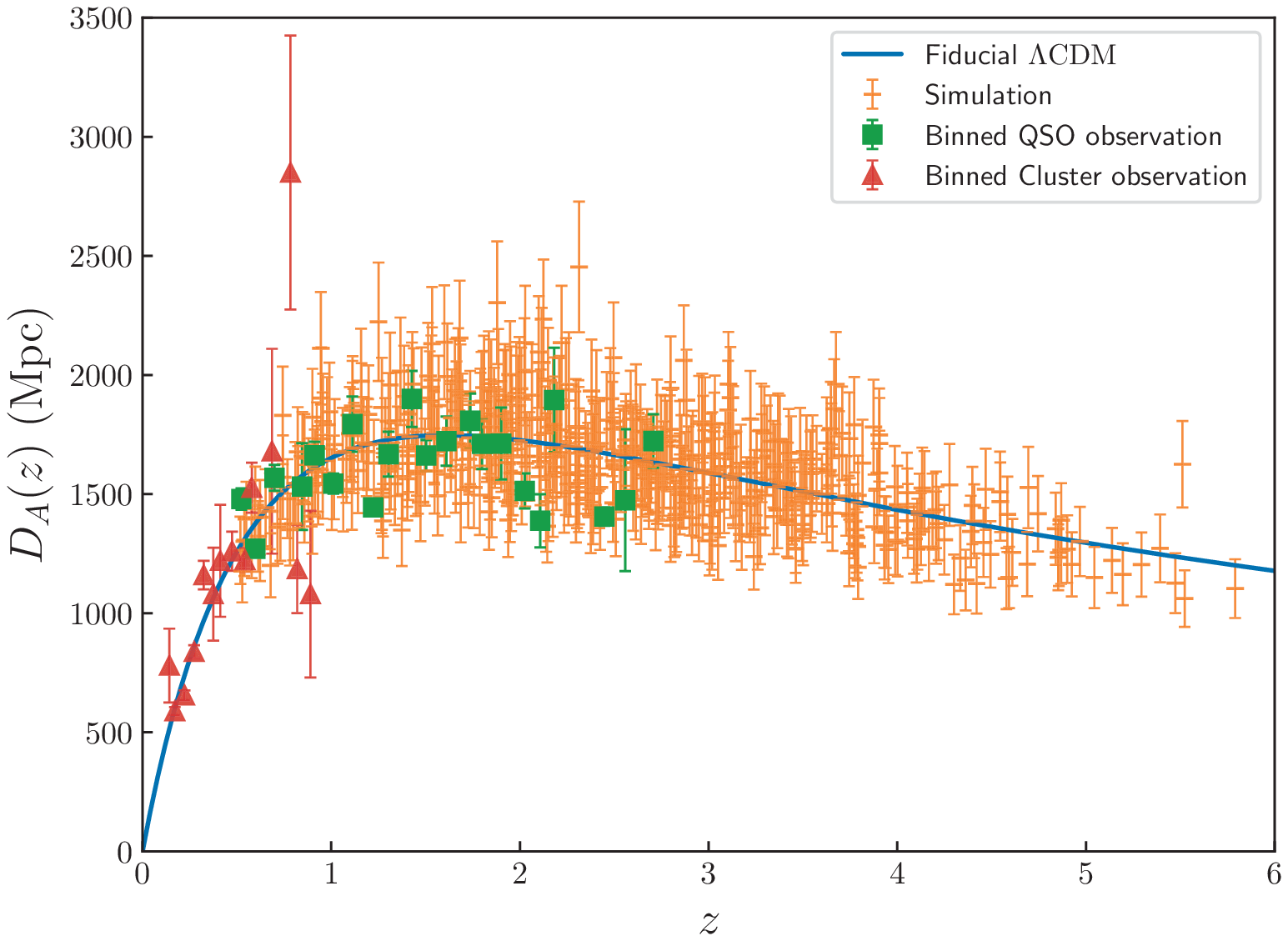} \includegraphics[scale=0.45]{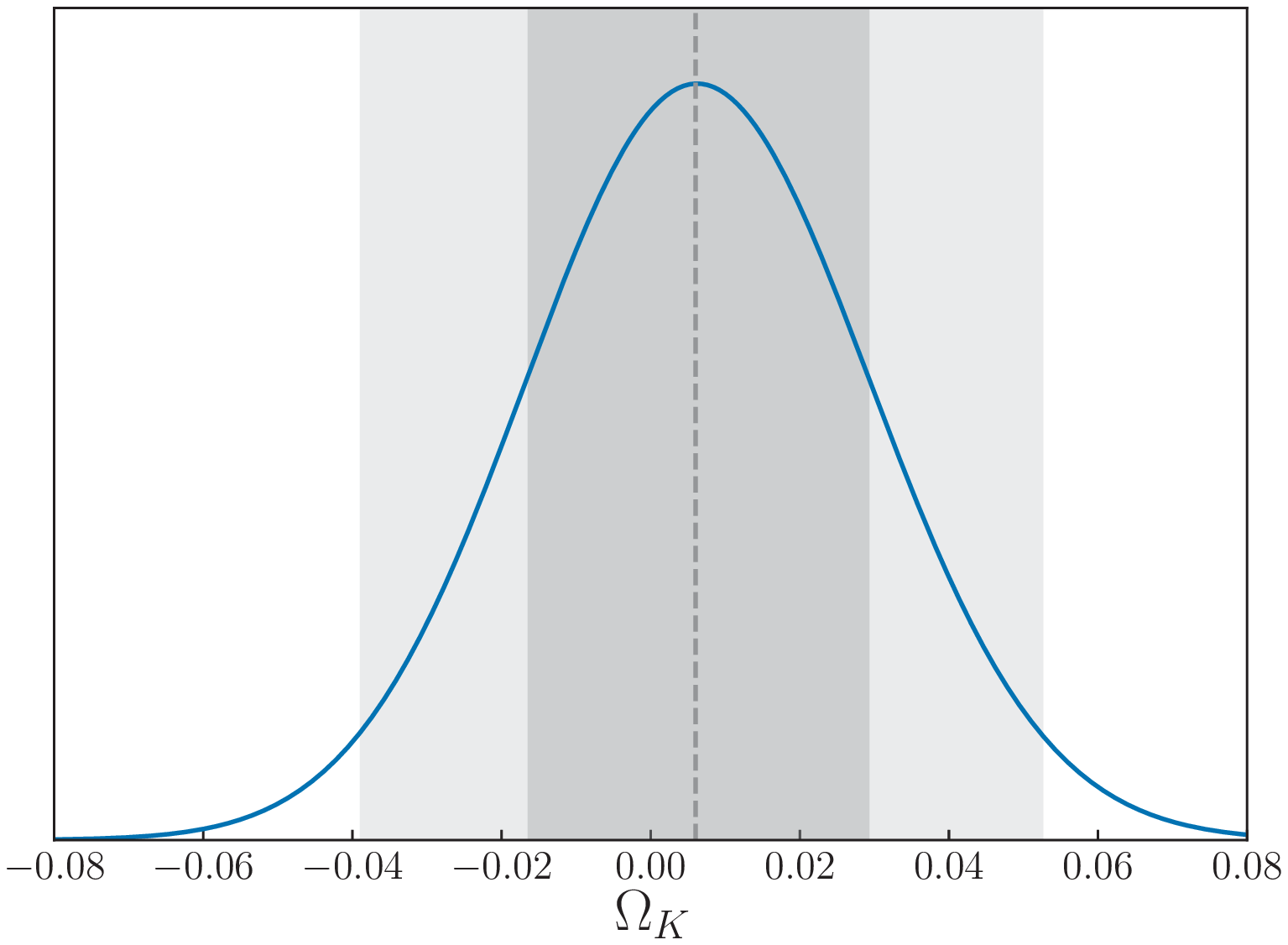}
\caption{ Simulated quasar data (left panel) and the corresponding
constraint on the cosmic curvature (right panel). Angular diameter
distances estimated from quasars as standard rulers (green squares)
and galaxy clusters (pink triangles) are also added for comparison.
}\label{fig2}
\end{figure*}

Next, we applied a cosmological-model-independent method to
calibrate the linear size of the compact structure in the ILQSOs.
For this purpose we used the well-measured angular diameter
distances from the BAO \citep{Blake12,Xu13,Samushia14} covering the
redshift range $0.35\leq z \leq 0.74$. In order to obtain $D_A(z)$
from BAO samples, one can use a reconstruction method based on
Gaussian Processes. Applying the redshift-selection criterion,
$\Delta z=|z_{QSO}-z_{BAO}|\leq 0.005$, we obtain certain
measurements of $D_A$, inferred from the BAO data, corresponding to
the quasar redshifts. Next we perform a similar fitting procedure,
so that the values of $D_A$ inferred from quasars match the BAO
ones. As a result, we obtain the following:
\begin{eqnarray}
&& l_m= 11.0^{+0.4}_{-0.4} \ \mathrm{pc}.  
\end{eqnarray}
In order to check the constraining power of quasars with this
corrected linear size, using the ``$\theta-z$" relation for the full
quasar sample, we get stringent constraints on the cosmic curvature
$\Omega_k=0.0\pm0.1$ corresponding to the priors of
$H_{0}=69.6\pm0.7$ km $\rm s^{-1}$ $\rm Mpc^{-1}$ and
$H_{0}=73.24\pm1.74$ km $\rm s^{-1}$ $\rm Mpc^{-1}$, respectively.
This is also presented in Fig.~1. Therefore, a universe with zero
curvature (spatially flat geometry) is strongly supported by the
available observations. This is the most unambiguous result of the
current dataset. Moreover, in the context of model-independent
testing of the cosmic curvature, quasars may achieve constraints
with higher precision at much higher redshifts, comparing with other
popular astrophysical probes including SN Ia. For instance, the
uncertainty of the measured $\Omega_k$ is at the level of
$\sigma_{\Omega_{k}}\simeq0.10$ with our quasar data, which is
significantly better than that of the Union2.1/JLA
\citep{Li16,Wei16} sample ($\sigma_{\Omega_{k}}\simeq0.20$). The
constraining power of the former is more obvious when the large size
difference between the samples is taken into consideration.

In order to test further the validity and efficiency of our method,
we performed Monte Carlo simulation to create mock ``$H(z) - z$''
and ``$\theta - z$'' data sets, with the concordance $\Lambda$CDM
chosen as a fiducial cosmology ($\Omega_m=0.30$,
$\Omega_\Lambda=0.70$). Following the simulation method proposed in
\citet{Yu16}, there are 20 mock $H(z) - z$ data points in the Hubble
parameter simulation, the redshifts of which are chosen equally
spaced in log$(1+z)$ covering the region $0.1\leq z \leq 3.0$. The
fractional uncertainty of these mock data was taken at a level of
1\%. This is a reasonable assumption concerning the ``$H(z)$''
measurements which will be achieved in future observations
\citep{Weinberg13}. The linear size of the compact structure in
radio quasars was characterized by a Gaussian distribution $l_m=11.0
\pm 0.4$ pc. The quasar simulation was carried out in the following
way: I) When calculating the sampling distribution (number density)
of quasars, we adopt the luminosity function constrained from the
combination of the SDSS and 2dF (2SLAQ) \citep{Richards05}. Note
that the bright and faint end slopes in this model agree very well
with those obtained other luminosity functions including
\citet{Hopkins07}. In each simulation, there are 500
intermediate-luminosity quasars covering the redshift range
$0.50\leq z \leq 6.00$ and 250 data points are located in the
redshifts of $0.50\leq z \leq 3.00$. II) We attribute the angular
size of compact structure ``$\theta$'' to each quasar, the
fractional statistical uncertainty of which is taken at a level of
5\%. This reasonable assumption of the ``$\theta$" measurements will
be realized from both current and future VLBI surveys based on
better uv-coverage \citep{Coppejans16}. Meanwhile, we have also
assumed an additional 5\% systematical uncertainty in the observed
angular sizes to account for intrinsic variance in the size of the
cores, which is well clear from many high angular resolution
observations of AGNs. III) This process is repeated 100 times for
each data set and then provides the distribution of determined
average $\Omega_k$, therefore the final results are unbiased. An
example of the simulations and the fitting results are shown in
Fig.\ref{fig2}. We demonstrate that with 250 well-observed radio
quasars, one can expect the zero cosmic curvature to be estimated
with the precision of $\Delta\Omega_k \sim 10^{-2}$.

Finally, there are several sources of systematics we do not consider
in this paper and which remain to be addressed in the future
analysis. On the one hand, it should be emphasized that the data
used in this paper have been obtained more than three decades ago
with the VLBI systems much less sensitive than modern ones. A few of
the sources have small total flux densities $S_{t}$, for which the
determination of angular size will be accompanied by large
measurement uncertainties (see Eq.~(2)). In order to take into
account such systematics, further progress in this direction can be
achieved by focusing on the total flux density data currently
available from newest VLBI observations
\footnote{http://astrogeo.org/vlbi\_images/}. On the other hand, for
sources with relatively small flux densities (i.e. compared with
correlated flux density values $S_c$ at long baselines), there could
arise a question of calibration uncertainties and, more importantly,
possible variability of the sources. Such effect might also
contribute to the scatter of the results in the context of
cosmological studies like in this paper. In order to minimize the
influence of a few sources with extremely large systematics, several
authors proposed to bin the data and to examine the change in median
angular size with redshift. This procedure can be traced back to the
original works by \citet{Gurvits99}. We also pin our hope on
multi-frequency VLBI observations of more compact radio quasars with
higher angular resolution based on better uv-coverage
\citep{Pushkarev15}, in which the dependency of linear size $l_m$ on
frequency $\nu$ should be taken into account, i.e., following the
conical jet model proposed by \citet{Blandford79}, the intrinsic
linear size at other frequencies can be modified as $l_m \propto
\nu^{-k}$ \citep{Cao18}. Therefore, the prospects for constraining
the cosmic curvature with quasars could be promising, with future
multi-frequency VLBI surveys comprising much more sources with
higher sensitivity and angular resolution. \\

\textbf{\ Acknowledgments }

This work was supported by the National Natural Science Foundation
of China under Grants Nos. 11690023, 11503001, and 11633001;
National Key R\&D Program of China No. 2017YFA0402600; Beijing
Talents Fund of Organization Department of Beijing Municipal
Committee of the CPC; the Fundamental Research Funds for the Central
Universities and Scientific Research Foundation of Beijing Normal
University; and the Opening Project of Key Laboratory of
Computational Astrophysics, National Astronomical Observatories,
Chinese Academy of Sciences. J.-Z. Qi was supported by the China
Postdoctoral Science Foundation (Grant No. 2017M620661). Y. Pan was
supported by CQ CSTC under grant No. cstc2015jcyjA00044, and CQ MEC
under grant No. KJ1500414. This research was also partly supported
by the Poland-China Scientific \& Technological Cooperation
Committee Project (No. 35-4). M. Biesiada was supported by the
Foreign Talent Introduction Project and the Special Fund Supporting
Introduction of Foreign Knowledge Project in China.

\end{document}